\pdfoutput=1 
\documentclass[10pt, a4paper]{article}
\usepackage{lrec2022} 
\usepackage{multibib}
\newcites{languageresource}{Language Resources}
\usepackage{graphicx}
\usepackage{tabularx}
\usepackage{soul}
\usepackage{amsmath}
\usepackage{booktabs}
\usepackage{multirow}
\usepackage{todonotes} 
\usepackage{pgfplots, pgfplotstable}
\pgfplotsset{compat=newest}
\usepgfplotslibrary{groupplots} 

\usepackage{titlesec}
\titleformat{\section}{\normalfont\large\bfseries\center}{\thesection.}{1em}{}
\titleformat{\subsection}{\normalfont\SmallTitleFont\bfseries\raggedright}{\thesubsection.}{1em}{}
\titleformat{\subsubsection}{\normalfont\normalsize\bfseries\raggedright}{\thesubsubsection.}{1em}{}
\renewcommand\thesection{\arabic{section}}
\renewcommand\thesubsection{\thesection.\arabic{subsection}}
\renewcommand\thesubsubsection{\thesubsection.\arabic{subsubsection}}

\usepackage{epstopdf}
\usepackage[utf8]{inputenc}

\usepackage{hyperref}
\usepackage{xstring}

\usepackage{color}

\title{A Study on the Ambiguity in Human Annotation of German Oral History Interviews for Perceived Emotion Recognition and Sentiment Analysis}
 
\name{\shortstack{Michael Gref$^{~1}$, Nike Matthiesen$^{2}$, Sreenivasa Hikkal Venugopala$^{1,3}$, Shalaka Satheesh$^{1,3}$\\
		Aswinkumar Vijayananth$^{1,3}$, Duc Bach Ha$^{1}$, Sven Behnke$^{1,4}$, Joachim Köhler$^{1}$}} 
\address{$^1$Fraunhofer Institute for Intelligent Analysis and Information Systems (IAIS), Germany\\
	$^2$Haus der Geschichte der Bundesrepublik Deutschland Foundation (HdG), Bonn, Germany\\
	$^3$University of Applied Sciences Bonn-Rhein-Sieg, Germany\\
	$^4$Autonomous Intelligent Systems (AIS), Computer Science Institute VI, Univ. of Bonn, Germany\\
	\{michael.gref, sreenivasa.hikkal.venugopala, shalaka.satheesh, aswinkumar.vijayananth,\\
	 duc.bach.ha, sven.behnke, joachim.koehler\}@iais.fraunhofer.de,  matthiesen@hdg.de}

\abstract{
For research in audiovisual interview archives often it is not only of interest what is said but also how. Sentiment analysis and emotion recognition can help capture, categorize and make these different facets searchable. In particular, for oral history archives, such indexing technologies can be of great interest. These technologies can help understand the role of emotions in historical remembering. However, humans often perceive sentiments and emotions ambiguously and subjectively. Moreover, oral history interviews have multi-layered levels of complex, sometimes contradictory, sometimes very subtle facets of emotions. Therefore, the question arises of the chance machines and humans have capturing and assigning these into predefined categories. This paper investigates the ambiguity in human perception of emotions and sentiment in German oral history interviews and the impact on machine learning systems. Our experiments reveal substantial differences in human perception for different emotions. Furthermore, we report from ongoing machine learning experiments with different modalities. We show that the human perceptual ambiguity and other challenges, such as class imbalance and lack of training data, currently limit the opportunities of these technologies for oral history archives. Nonetheless, our work uncovers promising observations and possibilities for further research. 
 \\ \newline  }

\begin{document}

\maketitleabstract

\section{Introduction}

Oral history archives are often large audiovisual data repositories composing numerous interviews of contemporary witnesses to historical events. Deep learning can help to make these archives more accessible quantitatively and qualitatively. A prominent example in recent years is automatic speech recognition for transcription of oral history interviews. 

However, many potential deep learning applications for oral history interviews are still untapped. In a recent survey, \newcite{Pessanha.2022.ComputOralHistoryOverview} discuss potential applications of computational technologies for oral history archives. Among other aspects, the authors point out the potential benefits of these technologies in understanding the changes in emotions during remembering, storytelling, and conversing. In conjunction with the transcriptions, researchers can better infer not only \textit{what} is being said but also \textit{how}. 
In a recent research project, we study sentiment analysis and emotion recognition for German oral history interviews as the foundation for such complex search and indexing approaches for archives.

Various challenges arise when transferring research results to real-world, "in-the-wild" applications. As with many AI-based approaches, suitable representative training data of adequate scale is one of the key challenges. 
For natural data sets, the current gold standard for data annotation is to use other people's perception of emotion as the learning target, cf.\ \cite{Schuller.2018.SERSurvey}. However, even the annotation is a significant challenge since it is ambiguous and subjective. Emotions actually felt by the recorded persons and the emotions perceived by annotators may differ---and human recognition rates usually do not exceed 90\,\%, cf.\ \newcite{Akcay.2020.SEROverview}.

We assumed this value to be an upper limit. 
\newcite{Dupre.2020.CompareVERSystems} compared the emotion recognition performance of humans and eight commercial systems using facial expression videos. The experiments show an overall human recognition accuracy of 72\,\% for the six basic Ekman emotions classes \cite{Ekman.1980.Emotions}.
A 48--62\,\% range in recognition accuracy was observed for eight different tested commercial systems. The authors found that the machines' accuracy was consistently lower for spontaneous expressions.

\newcite{Krumhuber.2020.FER14DatabaseHumanVsMachine} studied human and machine emotion recognition using fourteen different dynamic facial expressions data sets---nine with posed/acted and five with spontaneous emotions. 
For posed emotions, they report mean human recognition scores of roughly 60--80\,\%. However, for the spontaneous five corpora, the mean human scores are roughly 35--65\,\%.

Human-annotated training data is the crucial building block for emotion recognition and sentiment analysis machine learning systems. However, since the human perception of spontaneous emotions is ambiguous and subjective, we address this issue for oral history interviews in this paper. We investigate to what extent different persons perceive and annotate the emotions and sentiment of interviewees in oral history recordings differently, comparing annotations of three different persons on a German oral history data set. We believe this contributes to assessing the general capabilities of such approaches for oral history interviews. With initial experiments on three different modalities, we further study the influence of the annotation ambiguity and class imbalance for these tasks. We uncover challenges of sentiment analysis and emotion recognition for the oral history use case that need to be addressed in further research.

\section{The HdG \textit{Zeitzeugenportal}}
\label{section:zeitzeugenportal}

\textit{Zeitzeugenportal}\footnote{\url{https://www.zeitzeugen-portal.de}} (Portal of Oral History) is a German online service by \textit{Haus der Geschichte} (House of the History) Foundation (HdG) that offers a central collection of contemporary German oral history interviews. More than 8,000 clips from around 1,000 interviews can currently be found at \textit{Zeitzeugenportal}.

\subsection{\textit{Multi-Modal Mining for Oral History}}

In the research project \textit{Multi-Modal Mining of German Oral History Interviews for the Indexing of Audiovisual Cultural Heritage}, \textit{Fraunhofer IAIS} cooperates HdG to investigate complex search modalities for indexing oral history interviews. These audiovisual interviews illustrate the individual processing of history and demonstrate the multiperspectivity of personal views and experiences. 
Emotions are an important factor in the memory process, so automated analysis can help better understand the role emotions play in historical remembering. 

\subsection{The \textit{HdG} Oral History Data Set}

We selected 10 hours of German oral history interviews from the HdG \textit{Zeitzeugenportal} for our experiments. Our \textit{HdG data set} comprises 164 different interview videos of 147 interviewees.  
The selected interviews were recorded between 2010 and 2020. Thus, the selection is representative of the more recent videos on the portal---including both professional and non-professional speakers. In addition, we aimed to represent different emotions in the selection and create a heterogeneous data set in terms of age and gender.

For prepossessing the HdG data set for annotation, we apply our current automatic transcription system Fraunhofer IAIS Audio Mining \cite{Schmidt.2016.AudiominingCurrent} with our latest robust broadcast ASR model to create a raw ASR transcript, including punctuation. 
We use the ASR result to chunk the interviews into short segments at the longest speech pauses until we obtain segments of 30 seconds or less. 

Based on this segmentation, three employees at the \textit{Haus der Geschichte}, who have an academic background in history, annotated the emotions and sentiment segment-wise. At the same time, a reference transcription is obtained by correcting the ASR transcript. Details are presented in the following section. After the annotation, the HdG data set was split into speaker-independent training, development, and test subset for model training and evaluation, as presented in Table \ref{tab:data_set_overview}. 

The data set is not published and is only used in-house due to the General Data Protection Regulation and the personal rights of the interviewees.

\begin{table}[t]
	\centering
	\footnotesize
	\begin{tabular}{lrrr}
		\toprule
		\textbf{HdG Set}		& \textbf{Videos} 	&  \textbf{Segments} &   \textbf{Hours}  \\ 
		\midrule	
		Training		& 104	& 	1,863	& 	6.35 \\
		Development		& 27	& 	430		& 	1.44 \\
		Test		& 33	& 	471		& 	1.74 \\
		\bottomrule
	\end{tabular}
	\caption{Overview of HdG oral history data sets after annotation and split into speaker-independent subsets.}
	\label{tab:data_set_overview}
\end{table}

\section{Annotation of Perceived Emotions and Sentiment in Oral History}

As discussed, human perceptions of emotion and sentiment are often subjective, ambiguous and may differ greatly from the speaker-internal state. 
We use the phrase \textit{recognition of perceived emotion} to emphasize this issue and how machine learning systems are trained on such annotated data.  
Such systems merely reproduce human decoding competence that works with different levels to decode emotions: the verbal (text), para verbal (voice), and nonverbal (face) level, similar to the Shannon-Weaver Model of Communication \cite{Shannon.1949.TheoryOfCommunication}.  

\subsection{Emotion and Sentiment Annotation}

\pgfplotsset{multiplot/.style={ 
	title style={yshift=-3.mm},
	style={font=\footnotesize},
	ybar, 
	ymin=0, 	
	ymajorgrids, 
	ymax=2800,
	ybar=0pt,
	bar width=8pt,
	xmin=0,
	enlarge x limits = 0.2,
	width=0.38\textwidth,   
	height=0.2\textwidth,
	ylabel style={yshift=-1mm},
	xtick={0,1,2,3},
	xtick pos=left,
	ytick={0,1000,2000},
	ytick pos=left,
	nodes near coords, 
	nodes near coords style={font=\tiny,rotate=90, anchor=west},
	legend columns=3,
	legend style={font=\scriptsize}
	} 
}

\begin{figure*}[t]
	\begin{center}
	\vspace{-1mm} 
		\begin{tikzpicture}
			\begin{axis}[
				title={Happiness},
				multiplot,
				ylabel={Segment Count},
				]
				\addplot coordinates {
					(0,2312)	(1,279)	(2,145)	(3,28)
				};
				\addplot coordinates {
					(0,2115)	(1,479)	(2,159)	(3,11)
				};
				\addplot coordinates {
					(0,1937)	(1,617)	(2,174)	(3,36)
				};
			\end{axis}
		\end{tikzpicture} \vspace{-3mm}
		\begin{tikzpicture}
			\begin{axis}[
				title={Sadness},
				title style={yshift=0.5mm},
				multiplot,
				yticklabels={}
				]
				\addplot coordinates {
					(0,2139)	(1,367)	(2,178)	(3,80)
				};
				\addplot coordinates {
					(0,2219)	(1,490)	(2,53)	(3,2)
				};
				\addplot coordinates {
					(0,2017)	(1,522)	(2,161)	(3,64)						
				};
			\end{axis}
		\end{tikzpicture} 
	\begin{tikzpicture}
		\begin{axis}[
			title={Anger},
			multiplot,
			yticklabels={}
			]
			\addplot coordinates {
				(0,2519)	(1,180)	(2,64)	(3,1)
			};
			\addplot coordinates {
				(0,2233)	(1,492)	(2,37)	(3,2)
			};
			\addplot coordinates {
				(0,2372)	(1,348)	(2,38)	(3,6)
			};
			\legend{A, B, C}
		\end{axis}
	\end{tikzpicture} 
	\begin{tikzpicture}
	\begin{axis}[
		title={Surprise},
		multiplot,
		ylabel={Segment Count},
		]
		\addplot coordinates {
			(0,2682)	(1,49)	(2,28)	(3,5)
		};
		\addplot coordinates {
			(0,2167)	(1,523)	(2,73)	(3,1)
		};
		\addplot coordinates {
			(0,2594)	(1,152)	(2,13)	(3,5)
		};
	\end{axis}
	\end{tikzpicture} 
	\begin{tikzpicture}
		\begin{axis}[
			title={Disgust},
			multiplot,
			yticklabels={}
			]
			\addplot coordinates {
				(0,2648)	(1,70)	(2,29)	(3,17)
			};
			\addplot coordinates {
				(0,2468)	(1,262)	(2,33)	(3,1)
			};
			\addplot coordinates {
				(0,2377)	(1,283)	(2,88)	(3,16)				
			};
		\end{axis}
	\end{tikzpicture} 
	\begin{tikzpicture}
		\begin{axis}[
			title={Fear},
			title style={yshift=0.5mm},
			multiplot,
			yticklabels={}
			]
			\addplot coordinates {
				(0,2629)	(1,79)	(2,51)	(3,5)
			};
			\addplot coordinates {
				(0,2282)	(1,397)	(2,83)	(3,2)
			};
			\addplot coordinates {
				(0,2390)	(1,281)	(2,76)	(3,17)
			};
		\end{axis}
	\end{tikzpicture} \vspace{-4.5mm} 
	\caption{Histograms of the annotation scores for each emotion. Each color bar represents a different annotator.}
	\label{fig:histogram_emotions}
	\end{center}
\end{figure*}
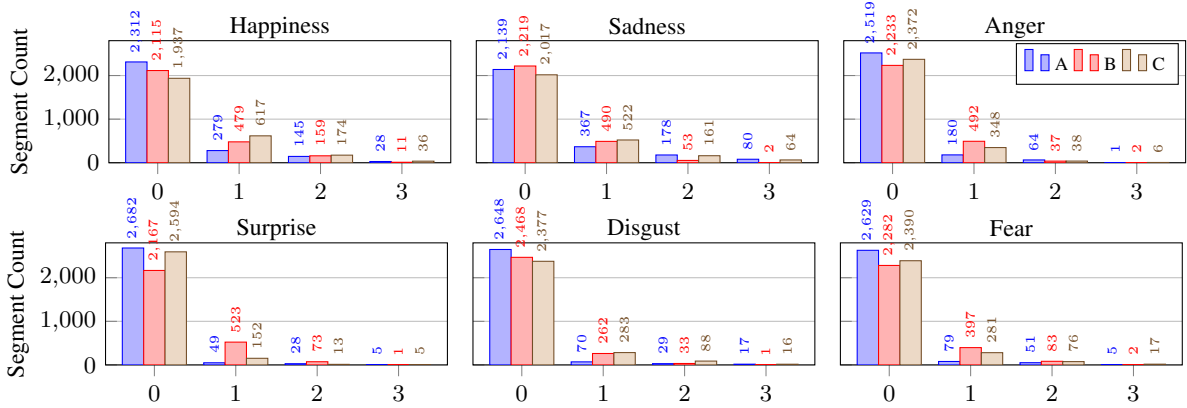

The annotation of emotion and sentiment is done in the same pass and on the same segments as correction of ASR transcripts. We use the six Ekman classes \cite{Ekman.1980.Emotions} for the raw emotion annotation: happiness, sadness, anger, surprise, disgust, and fear. Per segment, the three annotators assign a score on a 4-point Likert scale from 0 (no perception) to 3 (strong) for each of the six emotion classes following \citelanguageresource{Zadeh.2018.CmuMosei}. 
The annotation is done independently for each emotion class so that multiple emotions can appear in each segment to different degrees. Similar to the emotions, sentiment annotation is done on a Likert scale from -3 (very negative) to 3 (very positive).

\begin{figure}[t]
	\begin{center}
		\vspace{-1mm} 
		\begin{tikzpicture}
			\begin{axis}[
				style={font=\footnotesize},
				ybar, ymin=0, 	ymajorgrids, 
				ybar=0pt,
				bar width=6pt,
				enlarge x limits = 0.1,
				width=1.0\linewidth,   
				height=0.41\linewidth,
				xtick={-3,-2,-1,0,1,2,3},
				nodes near coords, 
				nodes near coords style={font=\tiny,rotate=90, anchor=west},
				legend style={font=\scriptsize},
				]
				\addplot coordinates {
					(-3,70)	(-2,255)	(-1,534)	(0,1374)	(1,356)	(2,147)	(3,28)
				};
				\addplot coordinates {
					(-3,28)	(-2,291)	(-1,667)	(0,817)		(1,763)	(2,183)	(3,15)
				};
				\addplot coordinates {
					(-3,25)	(-2,211)	(-1,744)	(0,1180)	(1,433)	(2,149)	(3,22)
				};
			\end{axis} 
		\end{tikzpicture}
		 \vspace{-4.5mm} 
		\caption{Sentiment annotation score histograms.}
		\label{fig:histogram_sentiment}
	\end{center}
\end{figure}
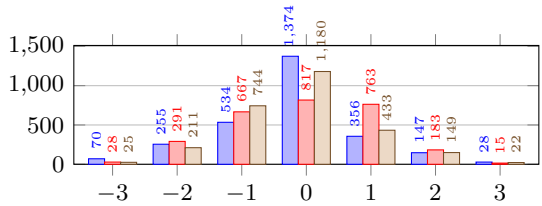

Figure \ref{fig:histogram_emotions} shows the distribution of annotated scores per emotion class for each of the three annotators. The distribution follows a pattern expected for natural, real-world data: A neutral score dominates for all emotions. With increasing score, the segment-count decreases. Although emotions play a decisive role in remembering, contemporary witnesses are often very composed when narrating. Therefore, a strong expression of emotions is rare. 
Happiness and sadness are the most represented emotions in our data, surprise and disgust are the weakest.

Figure \ref{fig:histogram_sentiment} presents the histogram of for the sentiment scores. As for emotions, the neutral score is most dominant. 
Unlike many other unstructured, real-world data sets, our data have negative sentiment more pronounced. This is likely due to the nature of the interviews: many interviews cover war and post-war experiences when Germany was divided into two states.

\subsection{Correlation Analysis of Annotation Pairs}

Table \ref{tab:class_correlation_between_transcribers} shows the class-wise relationship between the annotation for each pair of annotators in terms of correlation. We use the Spearman rank-order correlation coefficient measuring the monotony instead of a linear relationship between two annotators. Overall, the values for each annotator pair are in a similar range of values with no strong outliers. Therefore, we assume no fundamentally different understanding of the task or approach to annotation for any of the three annotators. 

\begin{table}[t]
	\centering
	\footnotesize
	\begin{tabular}{lrrr|r}
		\toprule
		& \multicolumn{3}{c}{\textbf{Transcriber Pairs}} &  \\ 
		\cmidrule(l{5pt}r{5pt}){2-4} 	
		\textbf{Class} 	& \textbf{A--B}	& \textbf{A--C}	& \textbf{B--C} & \textbf{Avg.}\\
		\midrule	
		sentiment	& 0.66	& 0.61	& 0.61 	& 	0.63 	\\
		happy		& 0.52	& 0.52	& 0.60 	&	0.55	\\
		sad			& 0.45	& 0.52	& 0.44 	&	0.47	\\
		anger		& 0.29	& 0.35	& 0.36 	&	0.33	\\
		surprise	& 0.14	& 0.26	& 0.19 	&	0.20	\\
		disgust		& 0.31	& 0.32	& 0.38 	&	0.34	\\
		fear		& 0.36	& 0.38	& 0.41 	&	0.38	\\
		\bottomrule
	\end{tabular}
	\caption{Spearman rank-order correlation coefficients between the annotated labels of two transcribers.}
	\label{tab:class_correlation_between_transcribers}
\end{table}

Sentiment has the strongest correlation among all classes. Thus, the annotators seemed to have comparatively the same perceptions regarding the sentiment. However, the correlation is just above moderate, with a mean value of 0.63, indicating some substantial differences between all three annotators.

Emotion is often considered more ambiguous and subjective than sentiment, as evidenced by the systematically lower correlation of these classes. Thus, there seem to be greater differences in perception or interpretation of emotions in our interviews. 
Happiness and sadness have the highest correlation coefficient with 0.55 and 0.47, respectively. Even if there is no consensus, we assume a fundamental agreement in a sufficient number of segments. 

The annotators seemed to have severely different perceptions for the other four emotion classes---with \textit{surprise} having the lowest correlation. Since even two humans seem to have only a conditionally identical perception for these emotions, we hypothesize that this ambiguity in annotation severely limits the recognition performance for oral history interviews---at least using these predefined classes. 


\subsection{Inter-Class Correlation Analysis}
\label{subsec:interclassdataanalysis}

\begin{figure}[t]
	\begin{center}
		\includegraphics[width=0.85\linewidth]{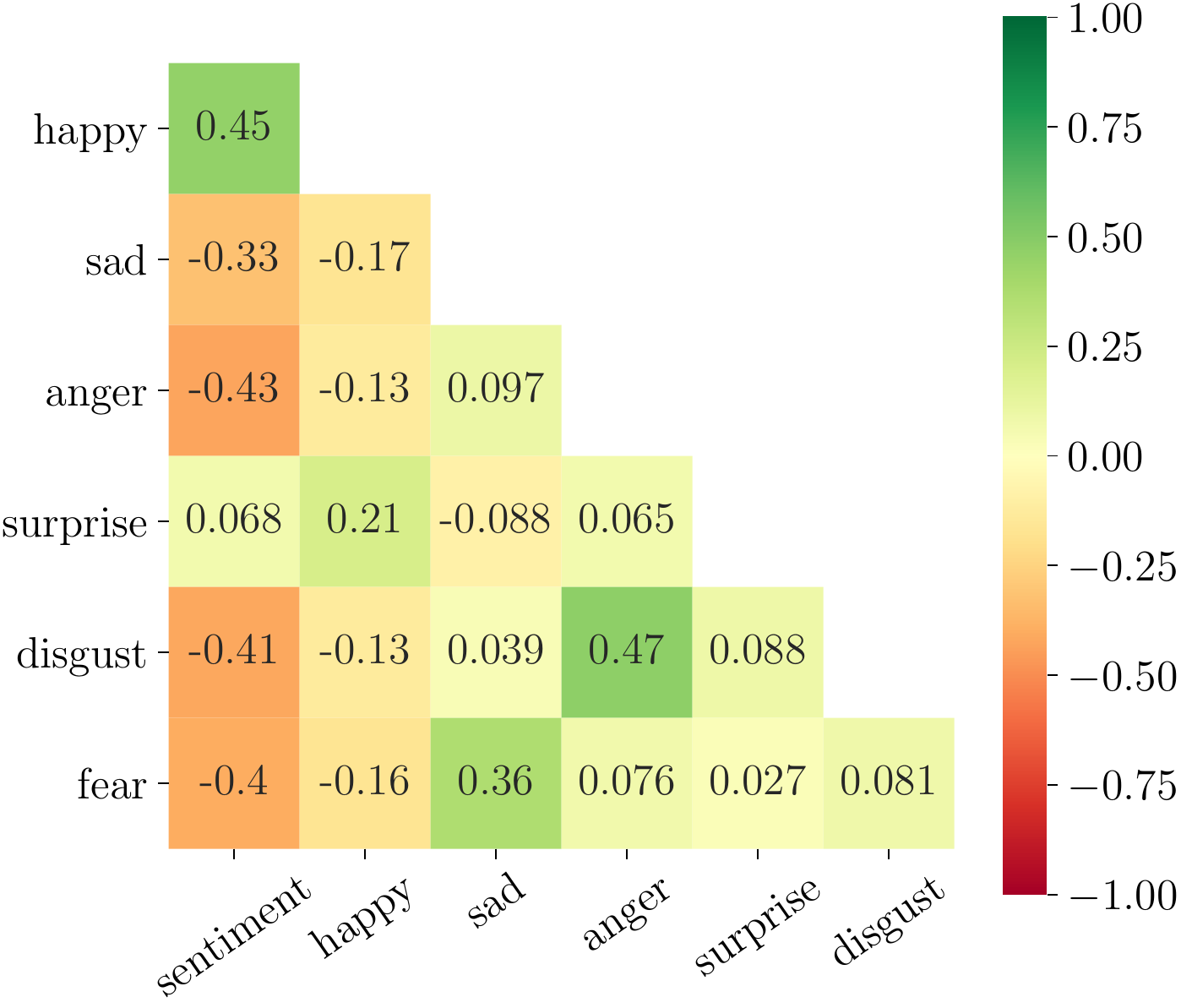}  \vspace{-3.5mm}
		\caption{Spearman correlation between the annotated average scores of the different classes.}
		\label{fig:emotion_correlation}
	\end{center}
\end{figure}

In a further correlation analysis, we investigate the co-occurrence of different emotions and sentiment. We combine the three annotations for this analysis by taking the arithmetic mean for each segment. A correlation matrix for the different classes is shown in Figure \ref{fig:emotion_correlation}. A moderate correlation exists between emotions and sentiment. In particular, happiness and a positive sentiment have a moderate correlation. An analogous correlation exists between negative sentiment and anger, disgust, and fear. 

In most cases, the correlation between the emotion classes is in the range of coincidence. Exceptions are disgust with anger (0.47) and fear with sadness (0.36). Thus, these class pairs seem to occur together more than just by coincidence and could be of interest for a detailed, qualitative analysis of the oral history interviews. We illuminate possible causes for these correlations by surveying the annotators.

\subsection{Qualitative Survey of Annotators}

In a qualitative survey, the annotators reported various challenges. One challenge is that the narrative structure of oral history interviews has different levels. Accordingly, emotions become visible in different ways, such as those that arise during remembering or reported emotional situations. In terms of the different emotions, the annotators agreed that the given Ekman classes are insufficient to reflect the complexity of emotions in oral history interviews. Nuances of emotions do not fit into the six categories. Therefore, the persons intuitively combined multiple emotions to represent more complex emotions, such as hate (disgust + anger), despair/helplessness (fear + sadness), scorn (happiness + disgust), and overwhelm (happiness + surprise) in the annotation. For example, overwhelm was identified as an important emotion in some interviews in which interviewees have talked about the Fall of the Berlin Wall. In combination, disgust and anger occurred more frequently in narratives reporting oppression or persecution.

\subsection{Mapping to Single-Label Data}
\label{subsec:single_label_train_dat}
The mean scores of the raw annotations were to corresponding classes for classification-system training. 
Our goal for the initial experiments was to keep it simple and compatible with common data sets to better understand the effects of ambiguity during training. Therefore, we aim to classify only the most prevalent emotion (single-label). For this, we use the arithmetic mean of the three annotations and proceed as follows: 

If the mean scores of all emotion classes are below 0.5, we assign this segment to \textit{neutral}. This aims to consider only emotion classes with trustworthy annotation, where at least two of three annotators have given a score of 1 ($ 0.\overline{6} $ on average), or at least one person has given a score of 2 and above. For non-neutral segments, we choose the class with the highest score. In exceptional cases, if two or more classes have the same score above the threshold, we mark them as \textit{ambiguous} and do not use them in the current training.

For the sentiment score $ s $, we apply a similar threshold and mapping: \textit{negative}, if $ s ~ \epsilon ~ [-3, -0.5] $; \textit{positive}, if $ s  ~ \epsilon ~  [0.5, 3] $; \textit{neutral}, if $ s  ~ \epsilon ~  (-0.5, 0.5)  $. 

Figure \ref{fig:piechart_single_label_emotions} shows the shares of each class in the entire HdG data set for both emotion and sentiment. Overall, the HdG data set is heavily imbalanced---both for the emotion and sentiment tasks.

\pgfplotsset{classdistplots/.style={ 
		title style={yshift=-2.5mm},
		style={font=\footnotesize},
		xbar, 
		xmin=0, 
		xmax=55,
		xmajorgrids,
		width=0.9\linewidth,   
		ytick distance=1,
		enlarge y limits={abs=0.2cm},
		nodes near coords, nodes near coords align={horizontal},
		bar width=7pt,
	} 
}

\begin{figure}[t]
	\begin{center} 	\vspace{-1.0mm} 
		\begin{tikzpicture}
			\begin{axis}[
				classdistplots,
				title={Emotions (Single Label)},
				symbolic y coords={surprise, disgust, fear, anger, ambiguous, sad, happy, neutral},
				/pgf/bar shift={0pt},  
				height=0.58\linewidth, 
				]
				\addplot coordinates {
					(42.0,neutral)
					(18.6,happy)
					(16.9,sad)
					(5.3,anger)
					(5.1,fear)
					(4.0,disgust)
					(2.6,surprise)
				};
				\addplot coordinates {
					(5.6,ambiguous)
				};
			\end{axis}
		\end{tikzpicture}
		\vfill
		\begin{tikzpicture}
		\begin{axis}[
			classdistplots,
			title={Sentiment (Single Label)},
			symbolic y coords={positive, negative, neutral},
			height=0.35\linewidth, 
			xlabel={Percentage of segments},
			]
			\addplot coordinates {
				(48.0,neutral)
				(31.2,negative)
				(20.8,positive)
			};
		\end{axis}
	\end{tikzpicture} 	\vspace{-4.5mm} 
		\caption{Percentages of emotion class and sentiment classes of whole data set after mapping to single-label.}
		\label{fig:piechart_single_label_emotions}
	\end{center}
\end{figure}
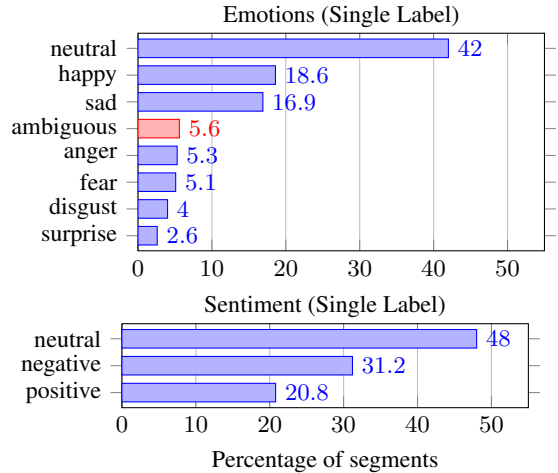

In the following three sections, we report results of initial, ongoing experiments and analysis with machine learning systems on our HdG data set. Experiments are performed on three different modalities: Speech, facial expressions, and text.

\section{Text-Based Sentiment and Emotion Recognition}

Sentiment analysis deals with mining opinions from an observer's point of view on content such as videos, speech, text, images. Opinion mining can be classified as polarity mining (sentiment polarity mining) and emotion mining.

\subsection{Related Work}
Sentiment and emotion analysis have been explored in various fields ranging from neuroscience, psychology, to human-machine interface. Comprehensive survey papers \cite{PangL07,Kao09,SalahABAQAA19,NorambuenaLV19,HemmatianS19,YadollahiSZ17} cover various approaches and methods on sentiment and emotion analysis tasks. 
Some of the comprehensive works on emotion analysis on text data are \newcite{MishneR06,RaoLLWQ14,AlmRS05,NeviarouskayaPI07}, and \newcite{GuptaGF13}. 
These works mainly consider that a document or a sentence consists of a single emotion. 
Only a few approaches deal with multi-label emotion analysis tasks \newcite{Bhowmick09,LiewT16,KhanpourC18}, and \newcite{Schoene20}.

\begin{table}[t]
	\centering
	\footnotesize
	\begin{tabular}{lrr}
		\toprule
		\textbf{}	& \textbf{Dev} 	&  \textbf{Test}  \\ 
		\midrule
		Sentiment	       	    & 70.7\,\% &	67.4\,\% \\
		Emotion	                & 34.6\,\%	& 33.9\,\%	 \\
		\bottomrule
	\end{tabular}
	\caption{Accuracy of the text-based sentiment and emotion models for the HdG test and development set.}
	\label{tab:senti_emoti_values}
\end{table} 

\subsection{Methodology and Implementation}
\label{subsec:sentiment_methods}
The pipeline used in our approach for sentiment analysis and emotion recognition starts with a BERT model to extract the embeddings from the tokenized text segments. We feed them into a classifier head consisting of two ReLU layers with a dropout layer in-between. We use the bert-base-german-cased pre-trained model\footnote{\url{https://huggingface.co/dbmdz/bert-base-german-cased}} as our base model. 

We apply a multi-stage training approach using the German part of the CMU-MOSEAS \citelanguageresource{Zadeh.2020.CmuMOSEAS} data set, mapped to single-label, and the HdG data set for fine-tuning in subsequent stages. 
In the first stage, we use the German CMU-MOSEAS subset comprising 1,000 videos with 480 distinct speakers distributed across 250 topics and 10,000 annotated sentences. 
In the second stage, we fine-tune the model using the HdG data set.
We use the raw ASR transcriptions from the pre-processing and not the human-corrected transcripts as inputs in the second stage. We aim to use the model as a subsequent analysis system after automatic transcription of large oral history data collections. On average, the HdG data set has a 16 to 17\,\% word error rate with our ASR system. 
To handle the class imbalance issue, we estimate the class weights using the \textit{compute class weights} function from the sklearn library that uses a heuristic inspired by logistic regression.

\subsection{Results and Inference}

\begin{figure}[t]
	\footnotesize
	\begin{center}
		\includegraphics[scale=0.3]{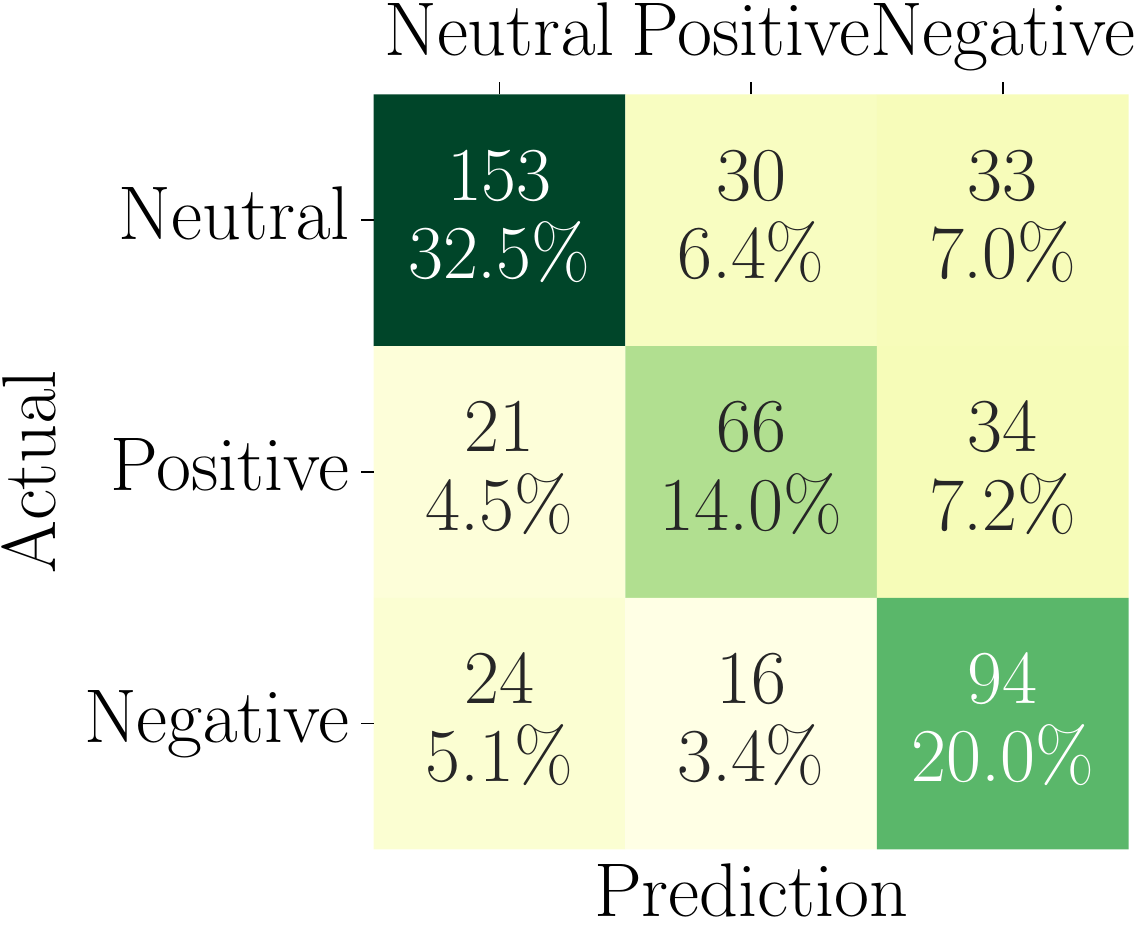} \vspace{2mm} \vfil
		\includegraphics[width=0.7\linewidth]{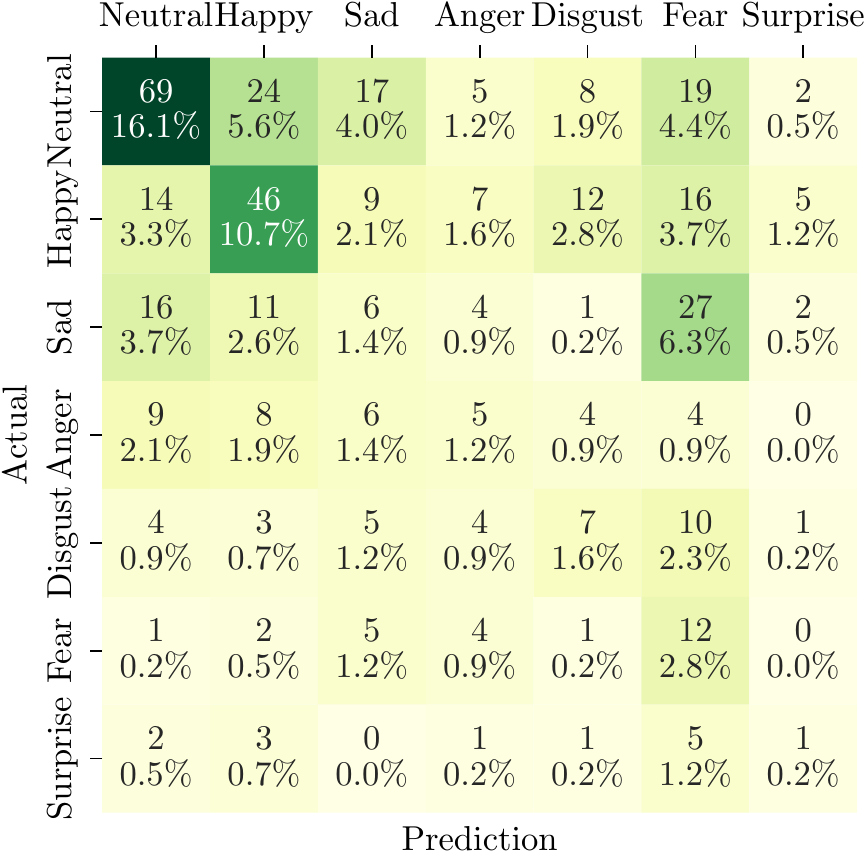}  \vspace{-3.5mm}
		\caption{Confusion matrix of the text-based sentiment analysis (top) and emotion recognition model (bottom) on HdG test.}
		\label{fig:text_cfmatrix}
	\end{center}
\end{figure}

The results of the sentiment and emotion classification are presented in Table \ref{tab:senti_emoti_values}. Our approach achieves a decent accuracy on the HdG sets for sentiment, however, only a low accuracy for the emotion recognition. The respective confusion matrices for both models are shown in Figure \ref{fig:text_cfmatrix}. For the sentiment model, we see only a few segments confused between all polarities. However, the emotion recognition model has only learned to distinguish between neutral and happiness.

Interestingly, we observe a slightly increased misclassification of the actual class sadness with fear. As our previous correlation analysis of the human annotations in Figure \ref{fig:emotion_correlation} has shown, these two classes have an increased correlation. The reason is that the annotators often intuitively combine these two emotions to express other emotions such as despair or helplessness. Therefore, this misclassification may not be a system failure but a limitation of the single-label approach.

Overall, recognizing emotions from oral history interviews on text alone seems very limited. Nevertheless, interesting observations emerge that deserve further research. This research should reveal whether the poor performance is due to the character of the interviews, the heavy class imbalance in training, or the modality text not conveying emotions appropriately without additional modalities. It might also be that the main reason is the ambiguity of the human annotation we observed on our data. We observe very similar results with the other modalities recognizing seven emotion classes. Therefore, we investigate these modalities with a subset of the classes in the following sections to uncover fundamental problems.

The text-based sentiment analysis works well on our unstructured, imbalanced oral history data. As the data analysis in Section \ref{subsec:interclassdataanalysis} indicated, there appeared to be a greater consensus among the three annotators on sentiment than emotion. This tendency seems to be confirmed by the experiment. In particular, we find it noteworthy that the classification works well given that we use raw, erroneous ASR transcripts as input to the model. 

\section{Speech Emotion Recognition}
Speech emotion recognition (SER) is a branch of affective computing that deals with identifying and recognizing the emotional content in speech. One of the significant challenges in this field is identifying appropriate features in the speech signal that best represent its emotional content.

\subsection{Related Work}

A detailed overview of SER is given, for example, by \newcite{Ayadi.2011.SurveySER}, \cite{Schuller.2018.SERSurvey}, and \cite{Akcay.2020.SEROverview}. Current approaches utilize convolutional neural network (CNN) or bidirectional recurrent neural network (biRNN) layers for SER---or combining both, such as \newcite{Dai.2019.LDFfromSpectrograms}. The proposed method represents emotion in speech in an end-to-end manner \cite{Zhang.2017.CNNforSER}. Furthermore, this method focuses on only four categories of emotion: \textit{anger, happy, sad}, and \textit{neutral}, which are identified as the most discriminatory ones. 

Further, \cite{Li.2019.DRNwithMultiheadAttention} propose a Diluted Residual Network (DRN) with multi-head self-attention. The authors employ Low-Level Descriptors (LLDs) extracted from the audio signal as input. 
\cite{Wang.2020.DualSequenceLSTM} propose a model consisting of two jointly trained LSTMs: each of these models is separately used to process MFCC features and Mel-spectrograms. Both models predict an output class (emotion) that is averaged to arrive at the result. Some of the currently used techniques also use transfer learning to boost the performance \cite{Akcay.2020.SEROverview}.

\subsection{Methodology and Implementation}

\begin{table}[t]
	\centering
	\footnotesize
	\begin{tabular}{lrr}
		\toprule
		& \multicolumn{2}{c}{\textbf{No.\ of Samples}}  \\ 
		\cmidrule(l{5pt}r{5pt}){2-3} 
		\textbf{Emotion} &\textbf{Original}	&	\textbf{Balanced}  \\ 
		\midrule
		Neutral      	& 9843		  & 3039             \\
		Happy         	& 3039		  & 3039             \\
		Sad	         	& 800		  & 2799             \\
		\bottomrule
	\end{tabular}
	\caption{Combined SER train data set before and after balancing with downsampling and data augmentation.}
	\label{tab:distribution-train-dataset-aug}
\end{table}

In this experiment, we train a hybrid model for SER, which combines traditional machine learning with deep learning. 
As for the text-based model, we utilize pre-trained models and multiple data sets to cope with the lack of training data. We apply a VGG-19 model pre-trained on the ImageNet data set and use log-Mel spectrograms treated as grayscale images as input features.

The pre-trained VGG-19 model is first fine-tuned on the HdG training set. Then we use a combined data set to extract the embeddings from the fine-tuned VGG-19. These embeddings are used as input for the SVM model. The combined data set contains the HdG train set, the German part of CMU-MOSEAS \citelanguageresource{Zadeh.2020.CmuMOSEAS}, CMU-MOSEI \citelanguageresource{Zadeh.2018.CmuMosei}, and \textit{Berlin Emotional Database} (Berlin EmoDB) \citelanguageresource{Felix.05.berlinemodb}. Except for CMU-MOSEI, an English data set, all other sets are German. Berlin EmoDB is the only set with acted emotions, whereas all other data sets have natural emotions. For data balancing, we apply data augmentation with 10 dB SNR additive white noise and downsampling the overrepresented. The distribution of the emotional classes in the combined train data set is presented in Table \ref{tab:distribution-train-dataset-aug}.

As already shown for the text modality, we have not achieved satisfactory recognition performance on our dataset so far with seven classes. Therefore, in this experiment, we only present results considering  \textit{happiness, sadness}, and \textit{neutral}. This aims to assess the problems of training better.

\subsection{Results and Inference}

\begin{figure}[t]
	\begin{center}
		\includegraphics[scale=0.30]{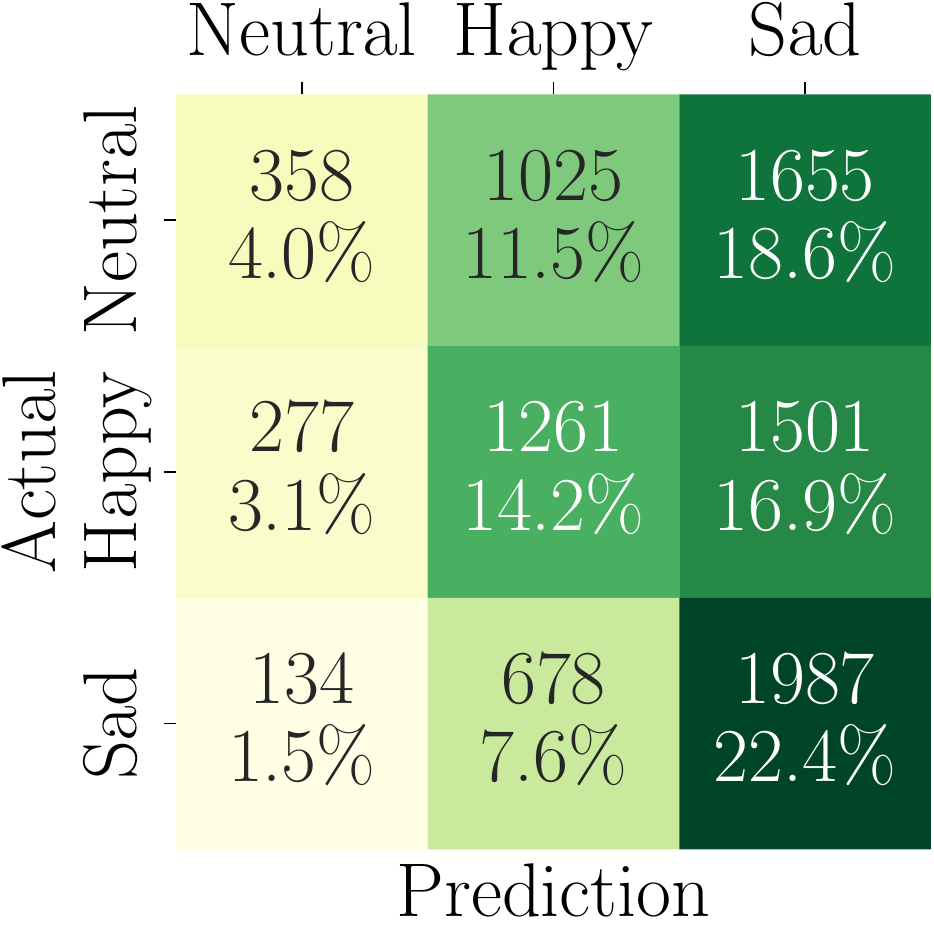}   \hspace{2mm} 
		\includegraphics[scale=0.30]{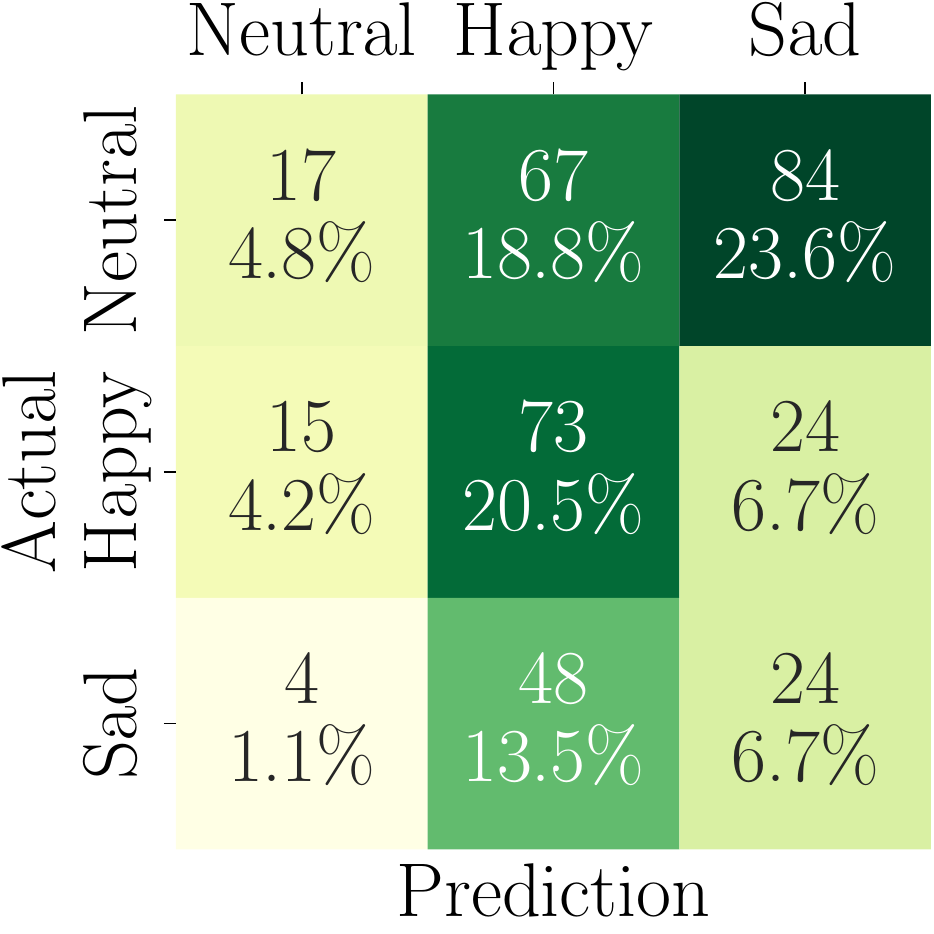} \vfill \vspace{1mm}
		\includegraphics[scale=0.28]{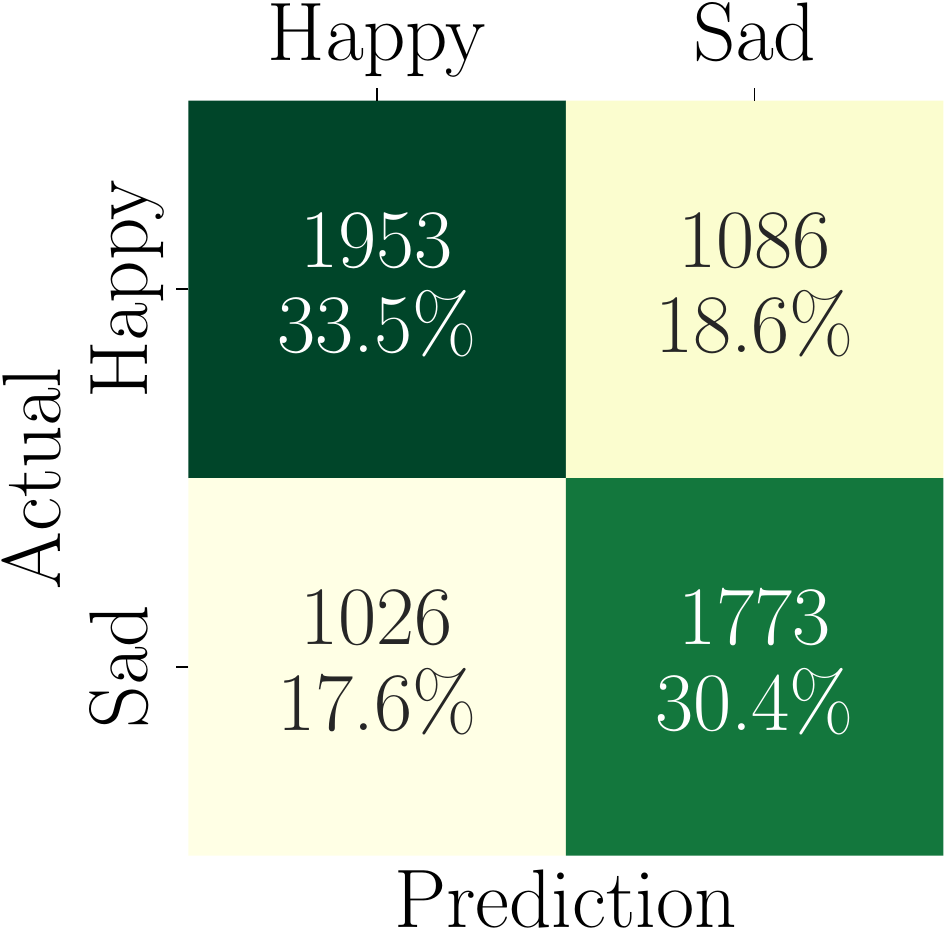} \hspace{2mm} 
		\includegraphics[scale=0.28]{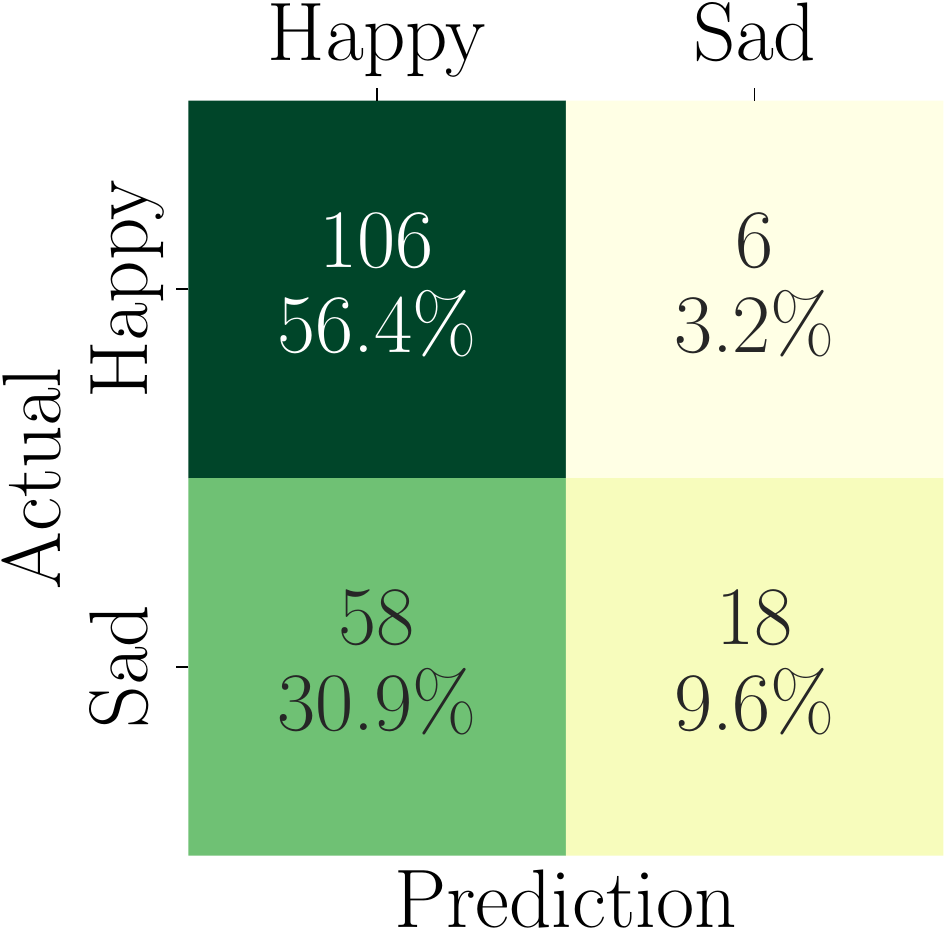}  \vspace{-3.5mm}
		\caption{Confusion matrices for 3 class (top row) and 2 class (bottom row) classification of the SER models on the combined training set (left) the HdG test set (right).}
		\label{fig:ser_cfmatrix}
	\end{center}
\end{figure}

The model yields a training accuracy 40.6\,\% and 32.0\,\% on the HdG test set for the three-class classification. The results are presented as a confusion matrix in the top row of Figure \ref{fig:ser_cfmatrix}. We observe that the neutral class is often confused with other emotions for both the training and test set. The results of the text modality already indicated this. However, it becomes more substantial in this experiment for the audio modality with three classes. We hypothesize that the neutral class cannot be sufficiently differentiated from subtle emotions in natural speech, leading to confusion in training. 

We conducted another experiment in which the neutral class is removed to investigate this issue further. For two-class classification (\textit{happy} and \textit{sad}) the accuracy improves to 63.8\,\% and 66.0\,\% for the training and test set, respectively. As shown at the bottom of Figure \ref{fig:ser_cfmatrix}, removing the neutral class results in a structural improvement. However, this does not lead to happiness and sadness being distinguished substantially better for the test set. The high accuracy is mainly attributed to the class imbalance towards happiness. Still, the system favors the happiness class over sadness.

A subjective evaluation of the HdG samples shows that it is challenging to differentiate emotions based on audio samples alone. Thus, we hypothesize that particular attention might have been paid to other modalities, presumably facial expressions, annotating the interviews.

\section{Facial Emotion Recognition}

\begin{figure*}[t]
	\begin{center}
		\includegraphics[scale=0.22]{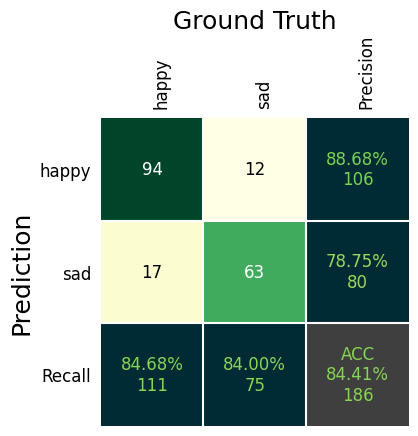} \hfil
		\includegraphics[scale=0.22]{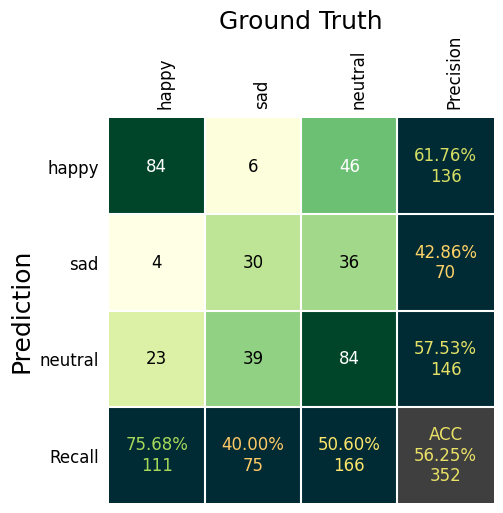} \hfil
		\includegraphics[scale=0.22]{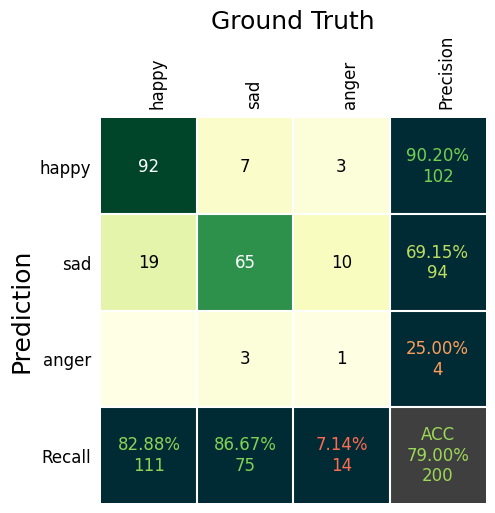} \hfil
		\includegraphics[scale=0.22]{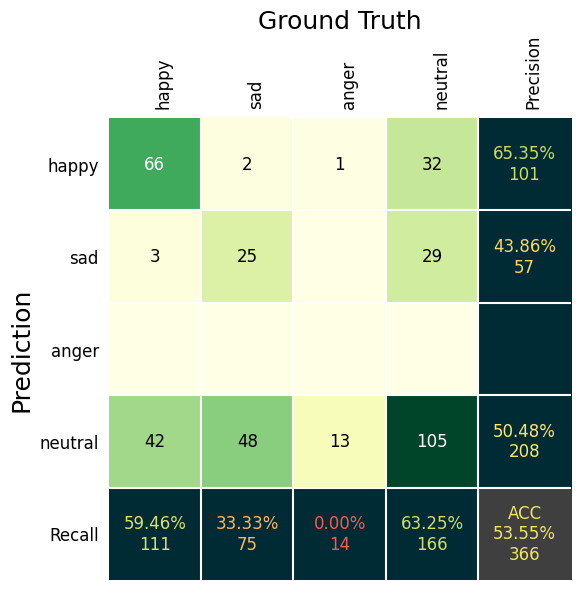}  \vspace{-3.5mm}
		\caption{Confusion matrices from results of the FER experiments.}
		\label{fig:fer_results}
	\end{center}
\end{figure*}

Facial emotion recognition (FER) is the task of recognizing human emotions from facial expressions. The immense number of visual clues present in the human face to identify underlying emotions makes FER an integral part of many emotion recognition systems.

\subsection{Related Work}

FER methods are categorized based on two dimensions: traditional feature- vs.\ representation-learning-based, and static vs.\ dynamic methods. Traditional feature learning-based FER methods rely on hand-crafted features. In contrast, representation-learning-based methods use a system such as a neural network to learn features from training data automatically. Dynamic methods utilize temporal relationships between frames of an input video while static ones treat every frame independently. Dynamic representation-learning approaches possess an inherent advantage and become potential candidates for further consideration. 

To perform the task at hand, we shortlisted \newcite{meng2019frame,kuo2018compact,gera2020affect,savchenko2021facial}, and \newcite{kuhnke2020two} based on factors such as performance on open-source FER data sets like CK+ \cite{lucey2010extended} and AFEW \cite{kossaifi2017afew}, depth of the neural network used (determines the minimum amount of data required for training), and reproducibility of results claimed by authors. Out of the five, Frame Attention Networks (FAN) \cite{meng2019frame} is chosen for its state-of-the-art accuracy on CK+ (99\,\%) and AFEW (51.18\,\%) data sets, and its simple yet effective construction. 

\subsection{Methodology}
The HdG videos are pre-processed using the d-lib based face detection and alignment modules to crop out and align faces. These sequences of images are used as inputs for the FAN. The FAN network architecture consists of a CNN-based feature embedding module (ResNet-18) and a subsequent frame attention network. 

\newcite{meng2019frame} offer three variants of FAN: baseline, self-attention, and the combination of self and relation attention. The authors report a slightly superior performance of the self-relation-attention variant over the other two. However, we currently use the baseline and self-attention variants due to their simple design, enabling us to better understand their work.

\subsection{Design of Experiments}
In addition to the challenges of emotion recognition in general and our HdG dataset in particular, we hypothesize an additional, specific challenge for FER. Most of the frames in a typical interview video carry faces with neutral or a subtle version of a particular emotion. This adds additional difficulty for any classifier to assign the correct label to the video---especially when \textit{neutral} is one of the possible target classes.

\begin{table}
	\centering
	\footnotesize
	\begin{tabular}{rrrrrr}
		\toprule
		\textbf{}	& \multicolumn{4}{c}{\textbf{HdG Training Samples}}  & \textbf{Test} \\ 
		\cmidrule(l{5pt}r{5pt}){2-5}
		\textbf{Exp} & \textbf{Happy} & \textbf{Sad} & \textbf{Anger} & \textbf{Neutral} & \textbf{Acc.} \\ 
		\midrule
			1 &	 318 & 316 & - 		& - 	& 84.4\% \\
			2 &	 318 & 316 & - 		& 316 	& 56.3\% \\
			3 &	 318 & 316 & 107	& - 	& 79.0\% \\
			4 &	 318 & 316 & 107 	& 787 	& 53.6\% \\ 	
		\bottomrule
	\end{tabular}
	\caption{Train data split for the four different FER experiments. The numbers for the emotion classes refer to the total number of segments of the applied HdG train set. For Experiment 2, neutral was reduced from 787 to 316 samples for class balancing.}
	\label{tab:train_split_fer}
\end{table}

We conducted four different experiments by training and evaluating the classifiers with different numbers and choices for the target emotion classes to study the effects of class-wise data set imbalance on the  model's performance. Table \ref{tab:train_split_fer} summarizes the experimental setup and the results. 

The first experiment was conducted with the already balanced pair of happy and sad classes, with an intent to study these classes' effect on the classifier's predictive performance. In the next experiment, the neutral class was included after under-sampling it to match the other two sizes. Whereas, for the third experiment, the under-represented anger class was added along with the "happy-sad" pair to understand the bias induced from the class imbalance. The final experiment was conducted with unchanged training data to evaluate the current state of the classifier's performance.

All experiments were conducted with both the baseline and self-attention variants of FAN. However, the results presented in the next section are limited to the baseline variant, which performed better in all of the conducted experiments. Models of both variants were pre-trained with Microsoft FER+ \cite{barsoum2016training} and AFEW \cite{kossaifi2017afew} data sets using transfer learning.

\subsection{Results and Inference}\label{sec_results_inference}

Figure \ref{fig:fer_results} shows the confusion matrices of the baseline variant of FAN in the four different experiments. The classifier exhibits a decent performance on the balanced pair of happy and sad classes in Experiment 1 with an overall accuracy of 84.4\,\%, proving its learning capacity. The high class-wise precision value indicates the model's discrimination capability on oral history interviews. 

However, the overall accuracy of the classifier drops significantly to 56.3\,\% with the addition of the neutral class in Experiment 2. This strongly indicates the neutral class's detrimental effect on the model's performance. The introduction of the neutral class possibly affects the classifier's sensitivity to identify subtle emotions which make most of the frames in a video. 

Unlike the neutral class, the under-represented anger class does not drastically reduce the classifier's accuracy. However, the model performs poorly on anger as it can correctly classify only one out of the fourteen test videos. This is certainly due to the insufficiency of anger in the training data. The model classifies most anger test videos as sad, presumably a relatively closer emotion to anger than happiness. 

Both discussed effects from inclusions of neutral and anger classes in Experiment 2 and 3 can be observed in a combined fashion from the results of Experiment 4. The over-represented neutral class hampers the classifier from correctly recognizing even a single test video from the anger class.

\section{Summary and Conclusion}

This work investigated the ambiguity in human perception of emotions and sentiment in German oral history interviews. Comparing the annotations of three persons using Ekman classes commonly used in emotion recognition revealed substantial differences in human perception. While the annotators in our experiment have a reasonably consistent understanding of the two most common emotions, \textit{happiness} and \textit{sadness}, we found very little correlation for other emotions.
Given the ambiguity of the human annotation using predefined emotions classes, we question whether practical learning for machines is even possible.

We further investigated co-occur of emotions in the annotation. An annotator survey revealed that Ekman classes were unanimously rated as insufficient for the complexity of multi-layered emotions in oral history interviews. The annotators intuitively combined different emotion classes to describe complex emotions not fitting in the predefined classes. This is reflected in an increased correlation of certain emotion classes, e.g., fear and sadness representing despair or helplessness. Hate was intuitively annotated as a combination of disgust and anger. 

We also reported results from initial emotion recognition experiments for facial expressions, speech, and text. A facial emotion recognition system for oral history revealed the system could differentiate happiness and sadness in our interviews. However, adding a neutral class results in the system not being able to differentiate between the subtle emotions and the neutral class. This issue and the combination of emotions described earlier are limited by single-label training. In future work with oral history, multi-label training should be considered to account for these aspects.

So far in our experiments, speech emotion recognition is behind facial emotion recognition. Even differentiating between happiness and sadness based on the voice appears challenging. For sentiment analysis based on raw ASR transcripts, on the other hand, we were able to achieve decent accuracy for our unstructured data. This is also consistent with the human perception, which was highest for sentiment between annotators in our experiments. 

In addition to the human ambiguity, other challenges currently limit the application of emotion recognition for oral history. In particular, we identified class imbalance and lack of representative training data as the current primary challenges. The application of pre-trained models, a combination of multiple natural data sets, and fine-tuning of models were essential in our work.

Overall, such indexing technologies for oral history archives seem to be quite limited so far. In oral history interviews, complex, subtle, and multi-layered emotions cannot yet be captured by our systems with the predefined, common classes. Perhaps fundamentally different approaches have to be chosen, e.g., limiting the indexing to recognizing specific patterns in human communication without interpreting them as emotions. 
However, users need to determine which patterns are relevant for their work in advance for meaningful application in archives. The results and observations of our work can provide initial impetus for this further research, which requires interdisciplinary collaboration between users of such archives and AI researchers.

\section{Acknowledgments}
The research project is funded by the German Federal Government Commissioner for Culture and Media.

\section{Bibliographical References}\label{reference}

\bibliographystyle{lrec2022-bib}
\bibliography{paper}

\section{Language Resource References}
\label{lr:ref}
\bibliographystylelanguageresource{lrec2022-bib}
\bibliographylanguageresource{languageresource}

\end{document}